\documentclass[11pt]{article}

\usepackage{chicagor}
\usepackage{times}
\newcommand{\commentout}[1]{}
\newtheorem{definition}{Definition}
\newcommand{\dfn}{\begin{definition}}
\newcommand{\edfn}{\bbox\end{definition}}
\newcommand{\bbox}{\vrule height7pt width4pt depth1pt}
\newcommand{\rarrow}{\rightarrow}

\begin{document}

\title{Computer Science and Game Theory: A Brief Survey}
\author{Joseph Y.\ Halpern%
\thanks{Supported in part by 
Work supported in part by NSF under grants
CTC-0208535 and ITR-0325453, by ONR under grant N00014-02-1-0455,
by the DoD Multidisciplinary University Research
Initiative (MURI) program administered by the ONR under
grants N00014-01-1-0795 and N00014-04-1-0725, and by AFOSR under grant
F49620-02-1-0101.  Thanks to Larry Blume, Christos Papadimitriou,
{\'E}va Tardos, and Moshe Tennenholtz for
useful comments.}\\  
   Cornell University\\
   Computer Science Department\\
   Ithaca, NY 14853\\
   halpern@cs.cornell.edu\\
   http://www.cs.cornell.edu/home/halpern
}
\maketitle

\section{Introduction}
There has been a remarkable increase in work at the interface of computer
science and game theory in the past decade.  Game theory forms a
significant component of some major computer science conferences 
(see, for example, \shortcite{EC05,aamas03}); leading computer scientists are
often invited to speak at major game theory conferences, such as the World
Congress on Game Theory 2000 and 2004.  In this article I survey some of
the main themes of work in the area, with a focus on the work in
computer science.  Given the length constraints, I
make no attempt at being comprehensive, especially since other surveys
are also available, including \shortcite{Hal38,Linial94,Papa01}, and a
comprehensive survey book will appear shortly \shortcite{NRTV07}.

The survey is organized as follows.
I look at the various roles of computational complexity 
in game theory in Section~\ref{sec:complexity}, including its use 
in modeling bounded rationality, its role in mechanism design, and 
the problem of computing Nash equilibria.  In Section~\ref{sec:anarchy},
I consider a game-theoretic problem that originated in the computer
science literature, but should be of interest to the game theory
community: computing the \emph{price of anarchy}, that is, the cost of using
decentralizing solution to a problem.  In Section~\ref{sec:gtdc} I
consider interactions between distributed computing and game theory.
I conclude in Section~\ref{sec:other} with a discussion of a few other
topics of interest.

\section{Complexity Considerations}\label{sec:complexity}
The influence of computer science in game theory has perhaps been most
strongly felt through complexity theory.  I consider some of the strands
of this research here.  There are a numerous basic texts on complexity
theory that the reader can consult for more background on notions like
NP-completeness and finite automata, including \shortcite{HU,Papa94}.

\paragraph{Bounded Rationality:} One way of capturing bounded
rationality is in terms of agents who have limited computational
power.  In economics, this line of research goes back to the work of
Neyman \citeyear{Ney85} and Rubinstein \citeyear{Rub85}, who focused on 
finitely repeated prisoner's dilemma.  
In $n$-round finitely repeated
prisoner's dilemma, there are $2^{2^n-1}$ strategies (since a strategy 
is a function from histories to $\{$cooperate, defect$\}$, and there
are clearly $2^n-1$ histories of length $< n$).  Finding a best response 
to a particular move can thus potentially be difficult.  Clearly people
do not find best responses by doing extensive computation.  Rather,
they typically rely on simple heuristics, such as ``Tit for Tat''
\shortcite{Axelrod}.   Such heuristics can often be captured by finite
automata; both Neyman and Rubinstein thus focus on finite automata
playing repeated prisoners dilemma.
Two computer scientists, Papadimitriou and Yannakakis \citeyear{PY94}, 
showed that if both players in an $n$-round 
prisoners dilemma 
are finite  automata with 
at least $2^{n-1}$ states, then the only equilibrium is the one
where they defect in every round.  This result says
that 
a finite automaton with exponentially many states
can compute best responses in prisoners dilemma.

We can then model bounded rationality by restricting the
number of states of the automaton.  Neyman \citeyear{Ney85} showed,
roughly speaking, that 
if the two players in $n$-round prisoner's dilemma are modeled by finite
automata with a number of states in the interval $[n^{1/k},n^k]$ for
some $k$, then collaboration can be approximated in equilibrium; more
precisely, if the payoff for (cooperate,cooperate) is (3,3)
there is an equilibrium in the repeated game where the average payoff
per round is greater than $3 - \frac{1}{k}$ for each player.  
Papadimitriou and Yannakakis \citeyear{PY94} sharpen this result by showing 
that if at least one of the players has fewer than 
$2^{c_\epsilon n}$ states, where $c_\epsilon =
\frac{\epsilon}{12(1+\epsilon)}$, then for sufficiently large $n$, 
then there
is an equilibrium where each player's average payoff per round is
greater than $3 - \epsilon$.   Thus, computational limitations can lead
to cooperation in prisoner's dilemma.  

There have been a number of other attempts to use
complexity-theoretic ideas from computer science to model bounded
rationality; see Rubinstein \citeyear{Rubinstein98} for some examples.
However, it seems that there is much more work to be done here.

\paragraph{Computing Nash Equilibrium:}  Nash \citeyear{Nash50} showed 
every finite game has a Nash equilibrium in mixed strategies.  But how
hard is it to actually find that equilibrium?  On the positive side, 
there are well known algorithms for computing Nash equilibrium, going
back to the classic Lemke-Howson \citeyear{LH64} algorithm, 
with a spate of recent improvements (see, for example, 
\shortcite{GW03,BSK03,PNS04}).  Moreover, for certain classes of games
(for example, symmetric games \shortcite{PR05}), there are known to be
polynomial-time 
algorithms.  On the negative side, many questions about Nash equilibrium
are known to be NP-hard.  For example, Gilboa and Zemel \citeyear{GZ89}
showed that, for a game presented in normal form, 
deciding whether
there exists a Nash equilibrium where each player gets a payoff of at
least $r$ is NP-complete.  (Interestingly, Gilboa and Zemel also show
that computing whether there exists a \emph{correlated} equilibrium
\shortcite{Aumann87} where each player gets a payoff of at least $r$ is
computable in polynomial time.  In general, questions regarding
correlated equilibrium seem easier than the analogous questions for Nash
equilibrium; see also \shortcite{Papa05,PR05} for further examples.)  Chu and
Halpern \citeyear{ChuHal98} prove similar NP-completeness results if the
game is represented in extensive form, even if all players have the same
payoffs (a situation which arises frequently in computer science
applications, where we can view the players as agents of some designer,
and take the payoffs to be the designer's payoffs).  Conitzer and
Sandholm \citeyear{CS03} give a compendium of hardness results for
various questions one can ask about Nash equilibria.  

Nevertheless, there is a sense in which it seems that the problem of
finding a Nash equilibrium is easier than typical NP-complete problems,
because every game is guaranteed to have a Nash equilibrium.  By way of
contrast, for a typical NP-complete problem like propositional
satisfiability, whether or not a propositional formula is satisfiable is
not known.  Using this observation, it can be shown that if finding a
Nash equilibrium is NP-complete, then NP = coNP.  Recent work has in a
sense has completely characterized the 
complexity of finding a Nash equilibrium in normal-form games:~it is a
\emph{PPPAD-complete} problem \cite{CD06,DGP06}.  PPAD stands for
``polynomial party argument (directed case)''; see \cite{Papa94a} for a
formal definition and examples of other PPAD problems.  It is believed
that PPAD-complete problems are not solvable in polynomial time, but are
simpler than NP-complete problems, although this remains an open
problem.  See \cite{Papa07} for an overview of this work.

\commentout{
A number of refinements of Nash equilibrium have been considered in the
literature, such as \emph{perfect equilibrium} \shortcite{Selten75},
\emph{sequential equilibrium} \shortcite{KW82}, 
and \emph{proper equilibrium} \shortcite{Myerson78}.  To the best of my
knowledge, the complexity of finding a more refined equilibrium has not
been studied in the literature.   However, it is easy to see 
that all the standard refinements can be
captured in the theory of \emph{real closed fields}
\shortcite{Tar1}---essentially, this is the first-order theory of the
real numbers.  More precisely, the question of whether there exists a
perfect (resp., sequential, proper) equilibrium within $\epsilon$ 
}
\paragraph{Algorithmic Mechanism Design:} The problem of mechanism
design is to design a game such that the agents playing the game,
motivated only by self-interest, achieve the designer's goals.
This problem has much in common with the standard computer science
problem of designing protocols that satisfy certain specifications
(for example, designing a distributed protocol that achieves
Byzantine agreement; see Section~\ref{sec:gtdc}). 
Work on  mechanism design has traditionally ignored
computational concerns.  But Kfir-Dahav, Monderer, and
Tennenholtz \citeyear{KMT00} show that, even in simple settings,
optimizing social welfare is NP-hard, so that perhaps the most common
approach to designing mechanisms, applying the Vickrey-Groves-Clarke (VCG)
procedure \shortcite{Clarke71,Groves73,Vickrey61}, is not going to work in
large systems.   We might hope that, even if we cannot compute an
optimal mechanism, we might be able to compute a reasonable
approximation to it.  However, as Nisan and Ronen \citeyear{NR00,NR01} show,
in general, replacing a VCG mechanism by an approximation does not
preserve truthfulness.  That is, even though truthfully revealing one's
type is an optimal strategy in a VCG mechanism, it may no longer be
optimal in an approximation.   Following Nisan and Ronen's work, 
there has been a spate of papers either describing computationally tractable
mechanisms or showing that no computationally
tractable mechanism exists for a number of problems, ranging from task
allocation 
\shortcite{AT01,NR01} to costsharing for multicast trees
\shortcite{FPS00} (where the problem is to share the cost of sending, for
example, a movie over a network among the agents who actually want the
movie) to finding low-cost paths between nodes in a network \shortcite{AT02}.

The problem that has attracted perhaps the most attention is
\emph{combinatorial auctions}, where bidders can bid on bundles of
items.  This becomes of particular interest in situations where the
value to a bidder of a bundle of goods cannot be determined by simply
summing the value of each good in isolation.  To take a simple example,
the value of a pair of shoes is much higher than that of the individual
shoes; perhaps more interestingly, an owner of
radio stations may value having a license in two adjacent cities more
than the sum of the individual licenses.  Combinatorial auctions are of
great interest in a variety of settings including spectrum auctions,
airport time slots (i.e., takeoff and landing slots), and industrial
procurement.   There are many complexity-theoretic issues related to
combinatorial auctions.  For a detailed discussion and references, see
\shortcite{CSS06}; I briefly discuss a few of the issues involved here.

Suppose that there are $n$ items being auctioned.  Simply for a bidder
to communicate her bids to the auctioneer can take, in general,
exponential time, since there are $2^n$ bundles.  
In many cases, we can identify a bid on a bundle with the bidder's
valuation of the bundle.  Thus, we can try to carefully 
design a bidding language in which a bidder can communicate her valuations
succinctly.  
Simple information-theoretic arguments can be used to show that, for
every bidding language, there will be valuations that will require
length at least $2^n$ to express in that language.  Thus, the best we
can hope for is to design a language that can represent the
``interesting'' bids succinctly.  
See \shortcite{Nisan06} for an overview of various bidding languages
and their expressive power.

Given bids from each of the bidders in a combinatorial auction, the
auctioneer would like to then determine the winners.  More precisely,
the auctioneer would like to allocate the $m$ items in an auction so as
to maximize his revenue.  This problem, called the \emph{winner determination
problem}, is NP-complete in general, even in relatively simple classes of
combinatorial auctions with only two bidders making rather restricted
bids.  Moreover, it is not even 
polynomial-time approximable, in the sense that there is no constant $d$
and polynomial-time algorithm such that the algorithm produces an
allocation that gives revenue that is at least $1/d$ of optimal.
On the other hand, there are algorithms that provably find a good
solution, seem to work well in practice,  and,  if they seem to
taking too long, can be terminated early, usually with a good feasible
solution in hand.  See \shortcite{LMS06} for an overview of the results in
this area.

In most mechanism design problems, computational complexity is seen as
the enemy.  There is one class of problems in which it may be a friend:
voting.  One problem with voting mechanisms is that of
\emph{manipulation} by voters.  That is, voters may be tempted to vote 
strategically rather than ranking the candidates according to their
true preferences, in the hope that the final outcome will be more
favorable.  This situation arises frequently in practice; in the 2000
election, American voters who preferred Nader to Gore to Bush were
encouraged to vote for Gore, rather than ``wasting'' a vote on Nader.
The classic Gibbard-Satterthwaite theorem \shortcite{Gibbard73,Satter75}
shows that, if there are at least three alternatives, then in any
nondictatorial voting scheme (i.e., one where it is \emph{not} the case
that one particular voter dictates the final outcome, irrespective 
of how the others vote), there are preferences under which an agent is
better off voting strategically.  The hope is that, by constructing the
voting mechanism appropriately, it may be computationally intractable to
find a manipulation that will be beneficial.  While finding
manipulations for majority voting (the candidate with the most votes
wins) is easy, there are well-known voting protocols for which
manipulation is hard in the presence of three or more candidates.
See \shortcite{CLS03} for a summary of results and further pointers to the
literature.  

\paragraph{Communication Complexity:} \emph{Communication complexity}
\cite{KN97} studies 
how much communication is needed for a set of $n$ agents to compute the
value of a function $f : \times_{i=1}^n \Theta_i \rarrow X$, where each
agent $i$ knows $\theta_i \in \Theta_i$.  To see the relevance of this
to economics, consider, for example, the problem of mechanism design.
Most mechanisms in the economics literature are designed so that agents
truthfully reveal their preferences (think of $\theta_i$ as
characterizing agent $i$'s preferences here).  However,
in some settings, revealing one's full preferences can require a
prohibitive amount of communication.  For example, in a combinatorial
auction of $m$ items, revealing one's full preferences may require
revealing what one would be willing to pay for each of the 
$2^m - 1$ possible bundles of items.  Even if $m=30$, this requires
revealing more than one billion numbers.  This leads to an obvious
question: how much communication is required by various mechanisms?
Nisan and Segal \citeyear{NS05} show that a standard approach for
conducting combinatorial auctions, where prices are listed, agents are
expected to make demands based on these prices, and then prices are
adjusted (according to some pre-specified rule) based on demand,
requires an exponential amount of communication for a certain class of
valuations.   This is among the first of preliminary steps to
understanding the communication complexity of mechanisms; the general
problem remains 
wide open.

\section{The Price of Anarchy}\label{sec:anarchy}
In a computer system, 
there are situations where we may have a choice between invoking a
centralized solution to a problem or a decentralized solution.
By ``centralized'' here, I mean that each agent in the system is told
exactly what to do and must do so; in the decentralized solution, each
agent tries to optimize his own selfish interests.  Of course,
centralization comes at a cost.  For one thing, there is a problem of
enforcement.  For another, centralized solutions tend to be more
vulnerable to failure.  On the other hand, a centralized solution may be
more socially beneficial.  How much more beneficial can it be?

Koutsoupias and Papadimitriou \citeyear{KP99} formalized this question
by comparing the ratio of the social welfare
of the centralized solution to that of the social welfare of the Nash
equilibrium with the worst social welfare (assuming that the
social welfare function is always 
positive).   They called this ratio the \emph{the price of anarchy}, and
proved a number of results regarding the price of anarchy for a
scheduling problem on parallel machines.
Since the original paper, the price of anarchy has been
studied in many settings, including traffic routing \shortcite{RT02},
facility location games (e.g., where is the best place to put a factory)
\shortcite{Vetta02}, and spectrum sharing (how should channels in a WiFi
network be assigned) \shortcite{HHLM04}.

  To give a sense of the results, consider the
traffic-routing context of Roughgarden and Tardos \citeyear{RT02}.  Suppose
that the travel time on a road increases in a known way with the
congestion on the road.  The goal is to minimize the average travel time
for all drivers.  Given a road network and a given traffic load,
a centralized solution would tell each driver which road to take.  For
example, there could be a rule that cars with odd-numbered license
plates take road 1, while those with even-numbered plates take road 2,
to minimize congestion on either road.   Roughgarden and Tardos show
that the price of anarchy is unbounded if the travel time can be a nonlinear
function of the congestion.
On the other hand, if it is linear, they show that the price of anarchy is
at most $4/3$.  

The price of anarchy is but one way of computing the ``cost'' of using a
Nash equilibrium.  Others have been considered in the computer science
literature.  For example, Tennenholtz \citeyear{Tenn02} compares the
\emph{safety level} of a game---the optimal amount that an agent can guarantee
himself, independent of what the other agents do---to what the agent
gets in a Nash equilibrium, and shows, for interesting classes of games,
including load-balancing games and first-price auctions, the ratio between
the safety level and the Nash equilibrium is bounded.  For example, in
the case of first-price auctions, it is bounded by the constant $e$.

\section{Game Theory and Distributed Computing}\label{sec:gtdc}

Distributed computing and game theory are interested in
much the same problems: dealing with systems where there are many
agents, facing uncertainty, and having possibly different goals.  
In practice, however, there has been a significant difference in 
emphasis in the two areas.  In distributed computing, the focus has been
on problems such as fault tolerance, asynchrony, scalability, and proving
correctness of algorithms; in game theory, the focus has been on
strategic concerns.  I discuss here some issues of common interest.%
\footnote{Much of the discussion in this section is taken from
\shortcite{Hal38}.}

To understand the relevance of fault tolerance and
asynchrony, consider the \emph{Byzantine agreement} problem,
a paradigmatic problem in the distributed systems literature.
In this problem, there are assumed to
be $n$ soldiers, up to $t$ of which may be faulty (the $t$ stands for {\em
traitor}); $n$ and $t$ are assumed to be common knowledge.
Each soldier starts with an initial preference, to either attack or
retreat. (More precisely, there are two types of nonfaulty
agents---those that prefer to attack, and those that prefer to retreat.)
We want a protocol that guarantees that 
(1) all {\em nonfaulty\/} soldiers reach the same decision, and
(2) if all the soldiers are nonfaulty and their initial preferences
are identical, then the final decision agrees with their initial
preferences.  
(The condition simply prevents the obvious trivial solutions,
where the soldiers attack no matter what, or retreat no matter what.)

This was introduced by Pease, Shostak, and Lamport \citeyear{PSL}, and
has been studied in detail since then; Chor and Dwork \citeyear{CD89},
Fischer \citeyear{Fisbyz}, and Linial \citeyear{Linial94} provide overviews.
Whether the Byzantine agreement problem is solvable depends in part on
what types of failures are considered, on whether the system is {\em
synchronous\/} or {\em asynchronous}, and on the ratio of $n$ to $t$.
Roughly speaking, a system is synchronous if there is a global clock and
agents move in lockstep; a ``step'' in the system 
corresponds to a tick of the clock. In an asynchronous system, there is
no global clock.  The agents in the system can run at arbitrary rates
relative to each other.  One step for agent 1 can correspond to an
arbitrary number of steps for agent 2 and vice versa.  Synchrony is an
implicit assumption in essentially all games.  Although it is certainly
possible to model games where player 2 has no idea how many moves player
1 has taken when player 2 is called upon to  move, it is not
typical to focus on the effects of synchrony (and its lack) in games.
On the other hand, in distributed systems, it is typically a major focus.

Suppose for now that we restrict to {\em crash failures}, where 
a faulty agent behaves according to the protocol, except that it might
crash at some point, after which it sends no messages.  In the round in
which an agent  fails, the agent may send only a subset of the messages
that it is supposed to send according to its protocol.  Further suppose
that the system is synchronous.  In this case, the following rather
simple protocol 
achieves Byzantine agreement: 
\begin{itemize}
\item In the first round, each agent tells every other agent its
initial preference.   
\item For rounds 2 to $t+1$, each agent tells every other agent
everything it has heard in the previous round.  (Thus, for example, in
round 3, agent 1 may tell agent 2 that it heard from agent 3 that
its initial preference was to attack, and that it (agent 3) heard from
agent 2 that its initial preference was to attack, and it heard from
agent 4 that its initial preferences was to retreat, and so on.  This
means that messages get exponentially long, but it is not difficult to
represent this information in a compact way so that the total
communication is polynomial in $n$, the number of agents.)
\item At the end of round $t+1$, if an agent has heard from any other
agent (including itself) that its initial preference was to attack, it
decides to attack; otherwise, it decides to retreat.
\end{itemize}

Why is this correct?  Clearly, if all agents are correct and want to
retreat (resp., attack), then the final decision will be to retreat
(resp., attack), since that is the 
only preference that agents hear about (recall that for now we
are considering only crash failures).  It
remains to show that if some agents prefer to attack and others to
retreat, then all the nonfaulty agents reach the same final decision.
So suppose that $i$ and $j$ are nonfaulty and $i$ decides to attack.
That means that 
$i$ heard that some agent's initial preference was to attack.  If it
heard this first at some round $t' < t+1$, then $i$ will forward this
message to $j$, who will receive it and thus also attack.  On the other
hand, suppose that $i$ heard it first at round $t+1$ in a message from
$i_{t+1}$. Thus, this message must be of the
form ``$i_t$ said at round $t$ that
\ldots that $i_2$ said at round 2 that $i_1$ said at round 1 that its
initial preference was to attack.''  Moreover, the agents $i_1,
\ldots, i_{t+1}$ must all be distinct.  Indeed, it is easy to see that
$i_k$ must crash in round $k$ before sending its message to $i$ (but after
sending its message to $i_{k+1}$), for $k = 1, \ldots, t$, for otherwise
$i$ must have gotten the message from $i_k$, contradicting the
assumption that $i$ first heard at round $t+1$ that some agent's initial
preference was to attack.  Since at
most $t$ agents can crash, it follows that $i_{t+1}$, the agent that
sent the message to $i$, is not faulty, and thus sends the message to
$j$.  Thus, $j$ also decides to attack.  A symmetric argument
shows that if $j$ decides to attack, then so does $i$.

It should be clear that the correctness of this protocol depends on both
the assumptions made: crash failures and synchrony.  Suppose instead that
{\em Byzantine\/} failures are allowed, so that faulty agents can deviate
in arbitrary ways from the protocol; they may ``lie'', send deceiving
messages, and collude to fool the nonfaulty agents in the most malicious
ways.  In this case, the protocol will not work at all.  In fact, it is
known that agreement can be reached in the presence of
Byzantine failures iff $t < n/3$, that is, iff fewer than a third of
the agents can be faulty \shortcite{PSL}.  The effect of asynchrony is even more
devastating: in an asynchronous system, it is impossible to reach
agreement using a deterministic protocol even if $t=1$ (so that there is at
most one failure) and only crash failures are allowed \shortcite{FLP}.  The
problem in the asynchronous setting is that if none of the agents have heard
from, say, agent 1, they have no way of knowing whether agent 1 is
faulty or just slow.  Interestingly, there are randomized algorithms
(i.e., behavior strategies) that achieve agreement with arbitrarily high
probability in an asynchronous setting \shortcite{BenOr,Rab}.  

Byzantine agreement can be viewed as a game where, at  each step, an
agent can either send a message or decide to attack or retreat.  It is
essentially a game between two teams, the
nonfaulty agents and the faulty agents, whose composition is unknown (at
least by the correct agents).  To model it as a game in the more
traditional sense, we could imagine that the nonfaulty agents are playing
against a new player, the ``adversary''.  One of adversary's moves is 
that of ``corrupting'' an agent: changing its type from ``nonfaulty'' to
``faulty''.   Once an agent is corrupted, what the adversary can do
depends on the failure type
being considered.  In the case of crash failures, the adversary can
decide which of a corrupted agent's  messages will be delivered in the
round in which the agent is corrupted; however, it cannot modify the
messages themselves.  
In the case of Byzantine failures, the adversary essentially gets to
make the moves for agents that have been corrupted; in particular, it can
send arbitrary messages.

Why has the distributed systems literature not considered strategic
behavior in this game?  
Crash failures are used to model hardware and software failures;
Byzantine failures are used to model random behavior on the part
of a system (for example, messages getting garbled in transit),
software errors, and malicious adversaries (for example, hackers).
With crash failures, it does not make sense to view the
adversary's behavior as strategic, since the adversary is not really
viewed as having strategic interests.  While it would certainly  
make sense, at least in principle, to consider the probability of
failure (i.e., the probability that the adversary corrupts an agent),
this approach has by and large been
avoided in the literature because it has proved difficult to
characterize the probability distribution of failures over time.
Computer components can perhaps be characterized as failing according to
an exponential distribution (see \shortcite{Bab87} for an analysis of
Byzantine agreement in such a setting), but crash failures can be
caused by things other than component failures (faulty software, for
example); these can be extremely difficult to characterize
probabilistically.   The problems are 
even worse when it comes to modeling random Byzantine behavior.

With malicious Byzantine behavior, it may 
well be reasonable to impute strategic behavior to agents (or to an
adversary controlling them).  However, it is often difficult to
characterize the payoffs of a malicious agent.
The goals of the agents may vary from that of simply trying to delay a
decision to that of causing disagreement.  It is not clear what the
appropriate payoffs should be for attaining these goals.
Thus, the distributed systems literature has chosen to focus instead on
algorithms that are guaranteed to satisfy the specification without
making  assumptions about the adversary's payoffs (or nature's
probabilities, in the case of crash failures).  

Recently, there has been some working adding strategic
concerns to standard problems in distributed computing (see, for
example, \shortcite{HT04}) as well as adding concerns of fault tolerance and
asynchrony to standard problems in game theory (see, for example,
\shortcite{Eliaz00,MT99,MT00} and the definitions in the next section).  
This seems to be an area that is ripe for further developments.
One such  development is the subject of the next section.

\section{Implementing Mediators}

The question of whether a problem in a multiagent system that can be
solved with a trusted mediator can be solved by just the agents in the
system, without the mediator, has attracted a great deal of attention in
both computer science (particularly in the cryptography community) and
game theory.  In cryptography, the focus on the problem has been on
\emph{secure multiparty computation}.  Here it is assumed that each
agent $i$ has some private information $x_i$.  Fix functions $f_1,
\ldots, f_n$.  The goal is have agent $i$ learn $f_i(x_1, \ldots, x_n)$
without learning anything about $x_j$ for $j \ne i$ beyond what is
revealed by the value of $f_i(x_1, \ldots, x_n)$.
With a trusted mediator, this is trivial: each agent $i$ just gives the
mediator its private value $x_i$; the mediator then sends each agent $i$
the value $f_i(x_1, \ldots, x_n)$.  Work on multiparty computation
\shortcite{GMW87,SRA81,yao:sc} provides conditions under which this can be
done.  In game theory, the focus has been on whether an equilibrium in a
game with a mediator can be implemented using what is called \emph{cheap
talk}---that is, just by players communicating among themselves
(cf.~\shortcite{Barany92,Bp03,Heller05,UV02,UV04}).  As suggested in the
previous section, the focus in the computer science literature
has been in doing multiparty computation in the presence of possibly
malicious adversaries, who do everything they can to subvert the
computation, while in the game theory literature, the focus has been on
strategic agents.  In recent work, Abraham et al.~\citeyear{ADGH06,ADH07}
considered deviations by both rational players, deviations
by both rational players, who have preferences and try to maximize them,
and players who can viewed as malicious, although it is perhaps better to
think of them as rational players whose utilities are
not known by the other players or mechanism designer.  I briefly sketch
their results here.%
\footnote{Much of the discussion in this section is taken from
\shortcite{ADH07}.}

The idea of tolerating deviations by coalitions of players goes back to
Aumann~\citeyear{Aumann59}; more recent refinements have been considered
by Moreno and Wooders~\citeyear{MW96}.  Aumann's definition is
essentially the following.

\dfn\label{def:1} $\vec{\sigma}$ is a \emph{$k$-resilient$'$
equilibrium} if, for all sets $C$ of players with $|C| \le k$, it is not
the case that there exists a strategy $\vec{\tau}$ such that
$u_i(\vec{\tau}_C,\vec{\sigma}_{-C}) > u_i(\vec{\sigma})$ for all $i
\in C$.
\edfn
As usual, the strategy $(\vec{\tau}_C,\vec{\sigma}_{-C})$ is the one
where each player $i \in C$ plays $\tau_i$ and each player $i \notin C$
plays $\sigma_i$.  As the prime notation suggests, this is not quite the 
definition we want to work with.  The trouble with this definition is
that it suggests that coalition members cannot communicate with each
other beyond agreeing on what strategy to use.  Perhaps surprisingly,
allowing communication can \emph{prevent} certain equilibria (see
\shortcite{ADH07} for an example).  Since we should expect 
coalition members to communicate, the following definition seems to
capture a more reasonable notion of resilient equilibrium.
Let the cheap-talk extension of a game $\Gamma$ be, roughly speaking, the
the game where players are allowed to communicate among themselves in
addition to performing the actions of $\Gamma$ and the payoffs are just as
in $\Gamma$.

\dfn\label{def:3} $\vec{\sigma}$ is a \emph{$k$-resilient 
equilibrium} in a game $\Gamma$ if $\vec{\sigma}$ is a 
$k$-resilient$'$ equilibrium in the cheap-talk extension of $\Gamma$
(where we identify the strategy $\sigma_i$ in the game $\Gamma$ with
the strategy in the cheap-talk game where player $i$ never sends any
messages beyond those sent according to $\sigma_i$).
\edfn

A standard assumption in game theory is that utilities are (commonly)
known; when we are given a game we are also given each player's
utility.When players make decision, they can take other players'
utilities into account.
However, in large systems, it seems almost
invariably the case that there will be some fraction of users who do
not respond to incentives the way we expect.  For example, in a
peer-to-peer network like Kazaa or Gnutella, it would seem that no
rational agent should share files. Whether or not you can get a file
depends only on whether other people share files; on the other hand, it
seems that there are disincentives for sharing (the possibility of
lawsuits, use of bandwidth, etc.). Nevertheless, people do share
files.  However, studies of the Gnutella network have shown almost
70 percent of users share no files and nearly 50 percent of
responses are from the top 1 percent of sharing hosts~\shortcite{AH00}.

One reason that people might not respond as we expect is that they have
utilities that are different from those we expect.   Alternatively,
the players may be irrational, or (if moves are made using a computer)
they may be playing using a faulty
computer and thus not able to make the move they would like, or they may
not understand how to get the computer to make the move they would like.
Whatever the reason, it seems important to
design strategies that tolerate such unanticipated behaviors, so that
the payoffs of the users with ``standard'' utilities do not get
affected by the nonstandard players using different strategies.
This can be viewed as a way of adding fault tolerance to equilibrium notions.

\dfn
A joint strategy $\vec{\sigma}$ is \emph{$t$-immune} if,
for all $T \subseteq N$ with $|T| \leq t$, all
joint strategies $\vec{\tau}$, and all $i \notin T$, we have
$u_i(\vec{\sigma}_{-T},\vec{\tau}_T) \ge u_i(\vec{\sigma})$.
\edfn

The notion of $t$-immunity and $k$-resilience address different
concerns.  For $t$ immunity, we consider the payoffs of the players
not in $C$; for resilience, we consider the payoffs of players in
$C$. It is natural to combine both notions.  Given a game $\Gamma$, let
$\Gamma^{T}_{\vec{\tau}}$ be the game that is identical to $\Gamma$
except that the players in $T$ are fixed to playing strategy
$\vec{\tau}$. 

\dfn
$\vec{\sigma}$ is a \emph{$(k,t)$-robust} equilibrium if
$\vec{\sigma}$ is $t$-immune and,
for all $T \subseteq N$ such that $|T| \leq t$ and all joint strategies
$\vec{\tau}$, $\vec{\sigma}_{-T}$ is a $k$-resilient strategy of
$\Gamma^{\vec{\tau}}_T$.
\edfn

To state 
the results of Abraham et al.~\citeyear{ADGH06,ADH07} on
implementing mediators, three games need to be considered:~an
\emph{underlying game} $\Gamma$, an extension $\Gamma_d$ of $\Gamma$
with a mediator, and a cheap-talk extension $\Gamma_{CT}$ of $\Gamma$.
Assume that $\Gamma$ is a \emph{normal-form Bayesian game}: each
player has a type from some type space with a known distribution over
types, and the utilities of the agents depend on the types and actions taken.
Roughly speaking, a cheap talk game \emph{implements} a game with a
mediator if it induces the same distribution over actions in the
underlying game, for each
type vector of the players. With this background, I can summarize the
results of Abraham et al.~\citeyear{ADGH06,ADH07}.  

\begin{itemize}
\item If $n > 3k + 3t$, a $(k,t)$-robust strategy $\vec{\sigma}$ with a
mediator can be implemented using cheap talk (that is, there is a
$(k,t)$-robust strategy $\vec{\sigma}'$ in a cheap talk game such that
$\vec{\sigma}$ and $\vec{\sigma}'$ induce the same distribution over
actions in the underlying game).  Moreover, the implementation
requires no knowledge of other agents' utilities, and the cheap talk
protocol has bounded running time
that does not depend on the utilities.

\item If $n \le 3k + 3t$ then, in general, mediators cannot be implemented
using cheap talk without  knowledge of other agents' utilities.
Moreover, even if other agents' utilities are
known, mediators cannot, in general, be implemented without having a
$(k+t)$-punishment strategy (that is, a strategy that, if used by all
but at most $k+t$ players, guarantees that every player gets a worse
outcome than they do with the equilibrium strategy) nor with bounded
running time. 

\item If $n > 2k+3t$, then mediators can be implemented using cheap talk
if there is a punishment strategy (and utilities are known) in
finite expected running time that does not depend on the utilities.

\item If $n \le 2k+3t$ then mediators cannot, in general, be implemented,
even if there is a punishment strategy and utilities are known.

\item If $n > 2k+2t$ and there are broadcast channels then, for all
$\epsilon$, mediators can be $\epsilon$-implemented (intuitively, 
there is an implementation where players get utility within $\epsilon$
of what they could get by deviating) 
using cheap talk, with
bounded expected running time that does not depend on the utilities.

\item If $n \le 2k+2t$ then mediators cannot, in general, be
$\epsilon$-implemented, even with broadcast channels.  Moreover, even
assuming cryptography and polynomially-bounded players, the expected
running time of an implementation must depend on the utility functions
of the players and $\epsilon$.

\item If $n > k+3t$ then, assuming cryptography and polynomially-bounded
players, mediators can be $\epsilon$-implemented using cheap talk,
but if $n \le 2k + 2t$, then the running time depends on the utilities
in the game and $\epsilon$. 

\item If $n \le k+3t$, then even assuming cryptography,
polynomially-bounded players, and a $(k+t)$-punishment strategy,
mediators cannot, in general, be $\epsilon$-implemented
using cheap talk.

\item If $n > k+t$ then, assuming cryptography, polynomially-bounded
players, and a public-key infrastructure (PKI), we can
$\epsilon$-implement a mediator.
\end{itemize}

The proof of these results makes heavy use of techniques from computer
science.  All the possibility results showing that mediators can be
implemented use techniques from secure multiparty computation.  The
results showing that that if $n \le 3k+3t$, then we cannot implement 
a mediator without knowing utilities, even if there is a punishment
strategy, uses the fact that Byzantine agreement cannot be reached if $t
< n/3$; the impossibility result for $n \le 2k + 3t$ also uses a
variant of Byzantine agreement.   

A related line of work considers implementing mediators assuming stronger
primitives (which cannot be implemented in computer networks); see
\shortcite{IML05,LMPS04} for details.

\section{Other Topics}\label{sec:other}

There are many more areas of interaction between computer science than I
have indicated in this brief survey.  I briefly mention a few others
here:

\begin{itemize}
\item \emph{Interactive epistemology:} Since the publication of Aumann's
\citeyear{Au} seminal paper, there has been a great deal of activity in
trying to understand the role of knowledge in games, and providing
epistemic analyses of solution concepts (see \shortcite{BB99} for a survey). 
In computer science, there has been a parallel literature applying
epistemic logic to reason about 
distributed computation.  One focus of this work has been on
characterizing the level of knowledge needed
to solve certain problems.  For example, to achieve Byzantine agreement
common knowledge among the nonfaulty agents of an initial value is
necessary and sufficient.  More generally, in a precise sense, common
knowledge is necessary  and sufficient for coordination.  
Another focus has been on defining logics that capture the reasoning of
resource-bounded agents.  This work has ranged from logics for reasoning
about awareness, a topic that has been explored in both computer science
and game theory (see, for example,
\cite{DLR98,FH,Hal34,HR05bnew,HMS03,MR94,MR99}) and logics for capturing
\emph{algorithmic 
knowledge}, an approach that takes seriously the assumption that agents
must explicitly compute what they know. See \shortcite{FHMV}
for an overview of the work in epistemic logic in computer science.

\item \emph{Network growth:} If we view networks as being built by
selfish players (who decide whether or not to build links), what will
the resulting network look like?  How does the growth of the network
affect its functionality? For example, how easily will influence spread
through the network?  How easy is it to route traffic?  
See \shortcite{FLMPS03,KKT03} for some 
recent computer science work in this burgeoning area.

\item \emph{Efficient representation of games:}  Game theory has
typically focused on ``small'' games, often 2- or 3-player games, that are 
easy to describe, such as prisoner's dilemma, in
order to understand subtleties regarding  basic issues such as rationality.
To the extent that game theory is used to tackle
larger, more practical problems, it will become important to find
efficient techniques for describing and analyzing games.  
By way of analogy, $2^{n} -1$ numbers are needed to describe a
probability distribution on a space characterized by $n$ binary random
variables.  For $n = 100$ (not an unreasonable number in 
practical situations), it is impossible to write down the probability
distribution in the obvious way, let alone do computations with it.
The same issues will surely arise in large games.  
Computer scientists use graphical approaches, such as \emph{Bayesian
networks}  and \emph{Markov networks} \shortcite{Pearl}, 
for representing
and manipulating probability measures on large spaces. Similar
techniques seem applicable to games; see, for example,
\shortcite{Koller01,LaMura00,KLS01}, and \shortcite{Kearns07} for a
recent overview.  Note that representation is also
an issue when we consider the complexity of problems such as computing
Nash or correlated equilibria.  The complexity of a problem is a
function of the size of the input, and the size of the input (which in
this case is a description of the game) depends on how the input is
represented.  

\item \emph{Learning in games:} There has been a great deal of work in
both computer science and game theory on learning to play well in
different settings (see \shortcite{FL98} for an overview of the work in
game theory).  One line of research in computer science has involved
learning to play optimally in a reinforcement learning setting, where an
agent interacts with an unknown (but fixed) environment.  The agent then
faces a fundamental tradeoff between \emph{exploration} and
\emph{exploitation}.  The question is how long it takes to learn to play
well (i.e., to get a reward within some fixed $\epsilon$ of optimal);
see \shortcite{BT02,KS98} for the current state of the art.
A related question is efficiently finding a strategy minimizes
\emph{regret}---that is, finding a strategy that is guaranteed to do not
much worse than the best strategy would have done in hindsight (that is,
even knowing what the opponent would have done).  See \shortcite{BM07}
for a recent overview of work on this problem.
\end{itemize}

\bibliographystyle{chicago}
\bibliography{z,joe,refs,game1}
\end{document}